\documentclass[proof]{WileyASNA-v1}

\articletype{PROCEEDING}%

\received{}
\revised{}
\accepted{}

\begin{document}

\title{Characterization of a C-RED One camera for astrophotonical applications}

\author[1]{Stella Vješnica*}
\author[1]{Eloy Hernandez}
\author[1]{Kalaga Madhav}
\author[1,2]{Martin M. Roth}

\authormark{Vješnica \textsc{et al}}

\address[1]{\orgdiv{innoFSPEC - Technical Section / Detector Systems}, \orgname{Leibniz-Institut f\"ur Astrophysik Potsdam (AIP)}, \orgaddress{\country{Germany}}}

\address[2]{\orgdiv{Institut f\"ur Physik und Astronomie}, \orgname{Universit\"at Potsdam}, \orgaddress{\country{Germany}}}

\corres{*Stella Vješnica, \email{svjesnica@aip.de}}

\presentaddress{Leibniz-Institut f\"ur Astrophysik Potsdam (AIP), An der Sternwarte 16, 14482 Potsdam, Germany}


\abstract{To better understand the impact of the avalanche gain applied in the detector technology and apply this technology in our in-house astrophotonic projects, we have characterized a C-RED One camera and produced a stable and reliable method for calculating the system gain at any desired avalanche gain setting. We observed that depending on how the system gain is obtained, multiplying the system gain times the avalanche gain may not accurately produce a conversion factor from electrons to ADUs. Since the acquisition of a photon transfer curve (PTC) was possible at different avalanche gain levels, several PTCs at low avalanche gain levels were acquired. Consequently, a linear fit was produced from the acquired system gain as a function of the avalanche gain setting. Through the linear fit, the effective system gain was calculated at any desired avalanche level. The effective system gain makes possible to accurately calculate the initial system gain without the ambiguity introduced by the non-linearity of the system. Besides, the impact of the avalanche gain on the dynamic range was also analyzed and showed a stable behaviour through the measured avalanche range. }


\jnlcitation{\cname{%
\author{Vješnica, S.}, \author{Hernandez, E.}, \author{Madhav, K.}, \author{Roth, M.M.}} (\cyear{2023}), 
\ctitle{Characterization of a C-RED One camera for astrophotonical applications}, \cjournal{Astronomische Nachrichten}, \cvol{2023;\textbf{???}}.}


\maketitle

\section{Introduction}

An avalanche photodiode (APD) works in principle as a pin-diode with an internal amplification through the avalanche effect, which is based on impact ionization produced by the photo-generated charge carriers. If the amplified carriers are electrons instead of holes, the denomination of eAPD is used \citep{Reisch}. 
Early attempts to produce focal plane arrays with the eAPD technology were conducted by ESO and Selex, nowadays Leonardo. The liquid phase epitaxy technology was used, but delivered improvable results. Afterwards, with the metal organic vapour phase epitaxy (MOVPE) technique, optimized heterostructure designs with a wide bandgap absorber region and a narrow bandgap gain region were possible. This allowed for a dramatic improvement in the cosmetic quality with 99.97$\%$ operable pixels at operating temperatures of 85K. This was followed by the development of the SAPHIRA ROIC for applications in the 1 to 2.5 $\mathrm{\mu m}$ wavelength range. With a 320x256 pixel format, 24 $\mathrm{\mu m}$ pitch, and 32 parallel video outputs operating at 5 MHz, readout times of 500 $\mathrm{\mu s}$ for a complete frame are possible, allowing for frame rates above 1 $\mathrm{kHz}$ with correlated double sampling \citep{Finger2014}.  The low read-out noise values at kilohertz frame rates confirm the revolutionary performance of these arrays with respect to the PICNIC or HAWAII technology \citep{Lanthermann2019}.  

There are already instruments that have upgraded the detector systems with this technology to achieve ground breaking results. As an example, the MIRC-X instrument, which is an upgrade of the Michigan Infrared Beam Combiner (MIRC), has reported read-out noise values of lower than 1 electron per frame per read \citep{Lanthermann2018}. MIRC-X is tied to the Center for High Angular Resolution Astronomy (CHARA) facility and in combination allowed imaging the surfaces of rapid rotators, interacting binary stars, and magnetically-active stars all for the first time ~\citep{Anugu2020}.  Another example of the success of eAPD is the upgrade for GRAVITY and the Very Large Telescope Interferometer (GRAVITY wide). In this case, read-out values of $\mathrm{0.15\text{ } e^{-}}$ rms were achieved \citep{Finger2019}. This instrument enabled astrometric measurements of stellar orbits around the supermassive black hole in the Galactic center~\citep{GRAVITY2018}, which led to the 2020 Nobel Prize in Physics. 

The French company First Light Imaging SAS (FLI) manufactured an autonomous system based on the SAPHIRA ROIC that offers an integrated cooling system and a vacuum regeneration system for off-the-shelf usage which is called the C-RED One camera \citep{Greffe2016}. The high-frame rate and low read-out noise va\-lues make this system optimal for both fringe tracking and wavefront sensing. At innoFSPEC in the Leibniz-Institute for Astrophysics Potsdam (AIP), there are several development projects in the field of astrophotonics that profit from these characteristics, i.e. the implementation of photonic beam combiners for astronomical interferometry in the infrared range to produce a stable setup and enable precise and higher resolution imaging ~\citep{Berger2001}; and  the combination of low-order adaptive optics (LOAO) and photonic lanterns technology, to improve the coupling efficiency of starlight into the fiber~\citep{Momen2021}.  

To better understand the impact of the avalanche gain applied in the detector technology and apply this technology in our in-house astrophotonic projects, we have characterized a C-RED One camera and produced a stable and reliable method for calculating the system gain at any desired avalanche gain setting. We observed that depending on how the system gain is obtained, multiplying the system gain times the avalanche gain may not accurately produce a conversion factor from electrons to ADUs. Since the acquisition of a photon transfer curve (PTC) was possible at different avalanche gain levels, several PTCs at low avalanche gain levels were acquired. Consequently, a linear fit was produced from the acquired system gains as a function of the avalanche gain settings. Through the linear fit, the effective system gain was calculated at any desired avalanche level. The effective system gain makes it possible to accurately calculate the initial system gain without the ambiguity introduced by the non-linearity of the system. Besides, the impact of the avalanche gain on the dynamic range was also analyzed and showed a stable behaviour through the measured range. 

A description of the characterized C-RED One camera is given in Sec. \ref{sec:c_red_one}. The detector characterization is explained in Sec. \ref{sec:characterization}, which include the dark current analysis, the system gain calculation through the PTC, the effective gain calculation, the readout noise calculations and sub-electrons RON results, and the dynamic range analysis. Chapter \ref{sec:conclusion} presents the conclusion.

\section{The C-RED One Camera}
\label{sec:c_red_one}

As mentioned in the Introduction, FLI created an off-the-shelf detector system which uses the SAPHIRA device, integrated in an appropriate read-out, vacuum and cryogenic coo\-ling system. The camera system is named the C-RED One and is de\-di\-cated for low-noise and fast imaging applications in the fields of near-IR wavefront sensing and fringe tracking. The camera was already used for the improvement of sensitivity in recent new interferometric instruments like the MIRC-X \citep{MIRC-X_2020}, as well as MYSTIC \citep{MYSTIC2018} and soon-to-be commissioned SILMARIL instrument (\cite{silmaril1_2022}; \cite{silmaril2_2022}).

The specifications of the camera are summarized in \hbox{Table \ref{tab1}}. The main characteristics include the use of a Mark 13 model 320x256 pixel frame SAPHIRA detector with embedded readout electronics system, placed in a sealed vacuum environment and cooled via pulse tube cooling at 80 K. The heat is dissipated by two cooling liquid hoses connected to a water chiller maintained at 20 $^{\circ}$C. Once cooled, the camera can acquire up to 3500 full frames per second and achieve the effective sub-electron readout noise at avalanche gains higher than 50. The frame rate can be even further increased, if one uses the sub-windowed mode.

\begin{center}
\begin{table}[htb]%
\centering
\caption{The C-RED One camera characteristics and performances, as specified by FLI.\label{tab1}}%
\tabcolsep=0pt%
\begin{tabular*}{20pc}{@{\extracolsep\fill}ll@{\extracolsep\fill}}
\toprule
\textbf{Main characteristics} & \textbf{Values} \\
\midrule
Detector format & 320$\times$256 pixels   \\
Pixel pitch & 24 $\mathrm{\mu}$m  \\
Wavelength & 0.8 - 2.5 $\mathrm{\mu}$m  \\
Operating temperature & 80 K  \\
Frame rate (full frame single read) & 3500 FPS  \\
Readout Noise (3500 FPS, gain $\sim 30$) & $< 1 \mathrm{e^-}$ \\
Dark current (gain $= 10$) & <80 $\mathrm{e^-/s}$  \\
Quantum Efficiency (QE) & $> 70 \%$  \\
Excess Noise Factor (ENF) & $< 1.25$  \\

\bottomrule
\end{tabular*}
\end{table}
\end{center}

Although the SAPHIRA detector can be used in \hbox{0.8 - 2.5 $\mathrm{\mu}$m} wavelength range, this camera is designed to include four long-wavelengths blocking filters that limit the sensitivity up to 1.75 $\mathrm{\mu}$m. This step is taken to filter out the unwanted background flux outside the J and H-band wavelength window, that would otherwise be amplified at low gain operation \citep{Gach2016}.

\section{Detector characterization}
\label{sec:characterization}

\begin{figure}[htb]
\centerline{\includegraphics[width=0.45\textwidth]{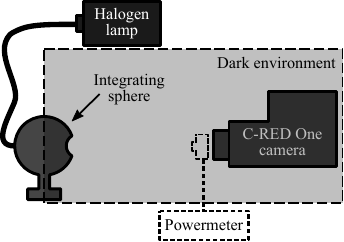}}
\caption{Schematic of the measurement setup used for acquisition of uniformly illuminated frames.\label{fig1}}
\end{figure}

For the characterization, flat field frames and dark frames were acquired. The flat field frames were obtained by illuminating the detector with a halogen light that passed through an integrating sphere. The integrating sphere has a diffuse reflectance coating that works in the effective spectral range of \hbox{350 - 2400 nm} with a reflectance of \hbox{96 - 98\%} in the \hbox{350 - 1000 nm} wavelength range and a \hbox{reflectance $> 90 \%$} up to 1750 nm. The light source  was connected to the integrating sphere through a set of liquid light guides. To avoid saturation and prevent damage to the pixels, the light flux at the camera opening was monitored with a power meter. In Figure \ref{fig1} a schematic of the complete setup is shown. 

The measurements were all conducted at an operating temperature of 80 K in four rounds of data acquisition.  Flat field frames with the following avalanche gain settings were used for the first acquisition round: 1, 2, 4, 6, 8, and 10. A value of 1 means that no avalanche gain is used. The second round included flat field frames with higher gain level: 20, 30, 40, and 50. For each avalanche gain setting, the exposure time was incremented until saturation was reached. For each step, 200 frames were recorded.  

The third and the fourth set of data consisted of multiple dark frames recorded at the minimal exposure time and different avalanche gain values (third set) or on the avalanche gain level 10 and different exposure times (fourth set). Dark frames (i.e. detector not exposed to light) were obtained with the camera cap mounted on, where the cap's outer surface was at room temperature. Here we emphasize how an insignificant difference in noise measurements at minimal exposure time was reported in \cite{Feautrier2022} between the configuration where camera was put in a cold environment, looking at a 80 K blackbody and the case where it observed a background at room temperature - demonstrating the efficiency of the long-wavelengths blocking filters mounted in front of the detector. 

\subsection{Dark current} \label{sec:DC}
\begin{figure}[htb]
\centerline{\includegraphics[width=0.47\textwidth]{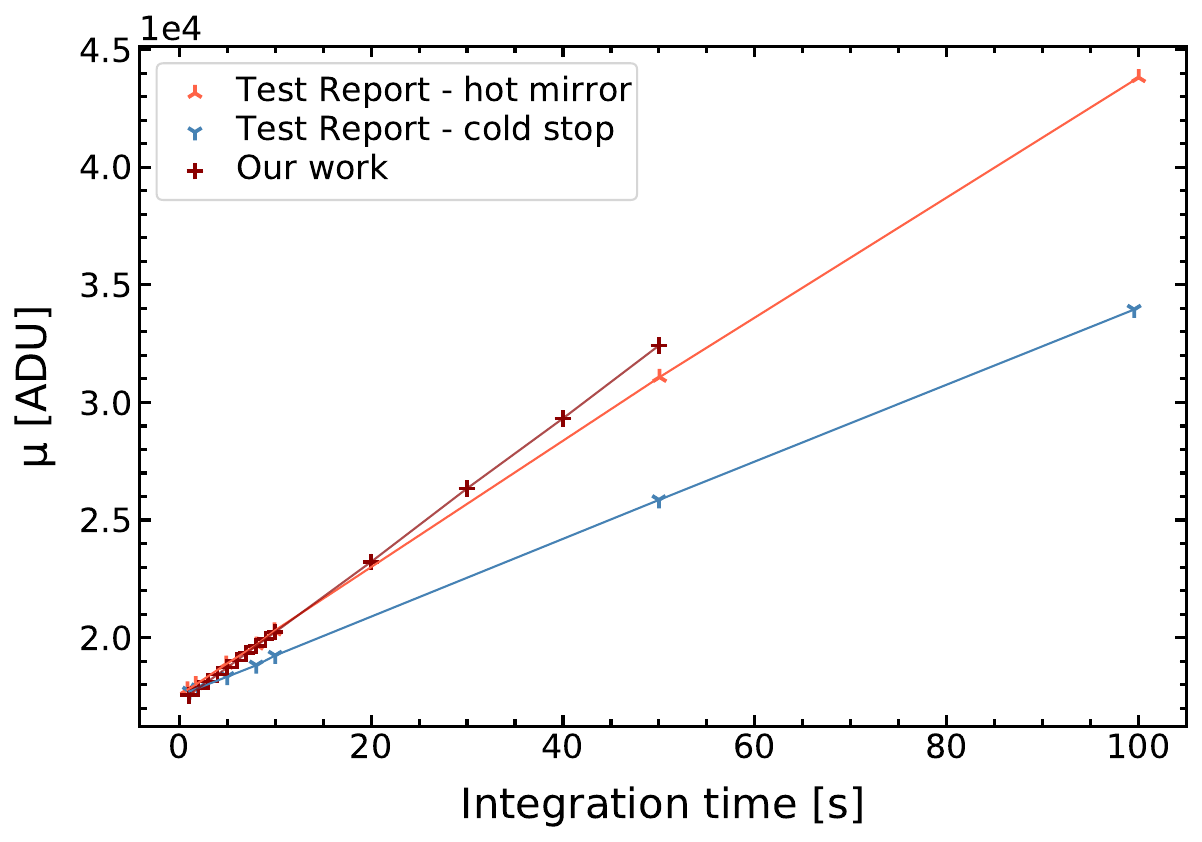}}
\caption{Measured dark current plus background radiation (darkred pluses) versus different integration times, in comparison to previously reported values from FLI (orange and blue signs).\label{fig0}}
\end{figure}

To examine the dark current of the C-RED One camera, dark frames were recorded at 14 different exposures in the range from 1 to 50 seconds. The frames were acquired while the camera was closed and therefore the closing cap at the room temperature was observed by the detector. We took 10 single, full frames at avalanche gain level 10 per exposure and calculated the average signal level of each pixel over the 10 identical exposures. We then averaged the 320$\times$256 points to determine the mean value across the detector. The same acquisition settings were used by the manufacturers, as reported in the test reports delivered with the camera. FLI acquired frames for two different setups with the camera viewing either: (a) an 80 K blackbody or (b) a hot mirror in front of the camera window. The data points from the FLI report are shown in the \hbox{Figure \ref{fig0}} together with the data we acquired. 

Using our calculated system gain for avalanche gain 10, the dark current we measure equals 76.98 e$^-$/pixel/sec. This current includes electrons from the $\sim 300$ K blackbody radiation of the lab and therefore the resulting trend is comparable to the case (b). Since the dark current is a temperature-dependent phenomenon, a slight discrepancy between the case (b) and our measurements may be explained by the difference in the room temperature at which measurements were performed.

\subsection{System gain} \label{sec:G_sys}

A fundamental property of the light signal is that the number of charge units fluctuates statistically and this statistic follow the Poisson distribution, where the variance of fluctuations is equal to the mean number of accumulated charges. This statement is true for the fundamental unit of quantization (electrons or photons). However, a detector system does not measure a signal in absolute physical units of electrons, but in relative analog-to-digital units. Therefore, a conversion factor to relate these units needs to be inferred from the slope of a plot of the temporal signal variance versus the mean signal. This conversion factor is better known as the system gain (G$\mathrm{_{sys}}$), while the relation itself is known as the Photon Transfer Curve (PTC) \citep{Janesick2001}. This method can provide a wealth of information regarding the dynamic range, gain and linearity of our system. 

\begin{figure}[htb]
	\centerline{\includegraphics[width=0.5\textwidth]{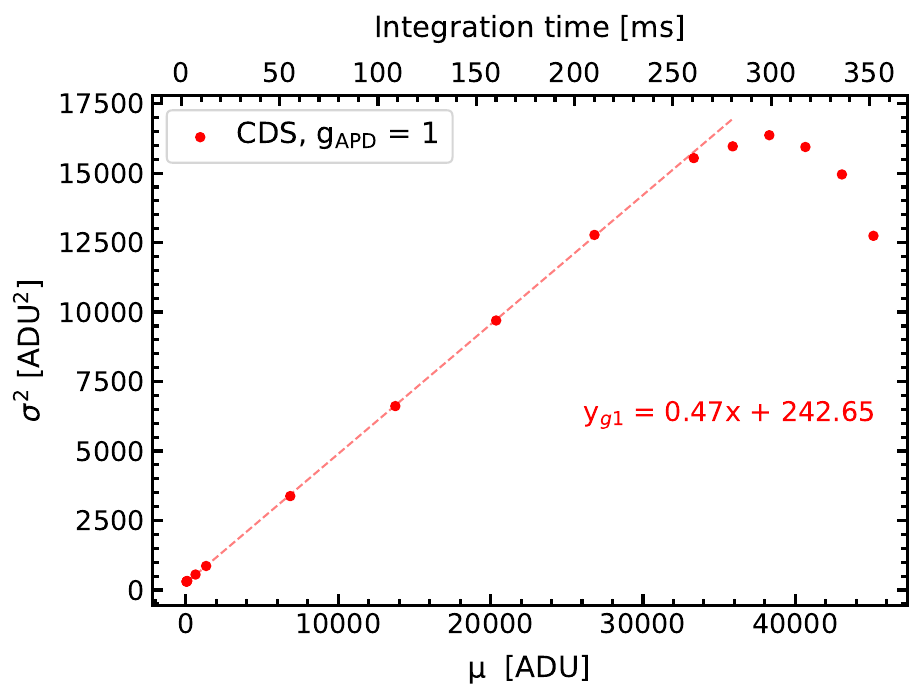}}
	\caption{PTC represented by the variance $\mathrm{\sigma}^2$ versus the average effective signal $\mathrm{\mu}$ of the correlated double sampling frames with the avalanche gain level set to 1. The linear relation is illustrated on the plot. Integration times are noted on the upper x-axis. \label{fig:gain_1}}
\end{figure}

From a sequence of 200 frames per integration time, a temporal signal variance was estimated at each pixel position. To represent each integration step with one variance value, temporal variance values from all the pixels contained in the flat field frame were averaged. With these averaged variance values and the averaged signal values we were able to construct a PTC diagram as illustrated in Figure~\ref{fig:gain_1}. This same method was applied for different avalanche gain settings in the first data set and will be of special interest in the analysis to follow in the next section. Currently, we focus only on the case where the avalanche gain (g$_{\mathrm{APD}}$) value is set to 1 (no avalanche process). From the slope of the PTC curve, a system gain G$\mathrm{_{sys}}$ is estimated to be \hbox{0.47 ADU/e$^-$} for frames obtained in Correlated Double Sampling (CDS) readout mode.

The G$\mathrm{_{sys}}$ displays slightly lower values from the ones found in different literature sources, where the analysis was also based on the PTC method. While \cite{Feautrier2022} report a higher value of G$\mathrm{_{sys}} =$ 0.76 ADU/e$^-$, \cite{Lanthermann2019} provides a value of 0.49 ADU/e$^-$, emphasizing possible variations in this parameter due to the non-linearity of the system and based on the portion of the PTC that is fitted. This point is very important as in the following section \ref{sec:eff_gain} we present a method that can mitigate this linearity problem and provide us with the universal tool of finding the appropriate conversion factor for different avalanche gain and flux levels.

\subsection{Effective gain calculation} \label{sec:eff_gain}

The basic system gain describes the conversion between charges and digital units, without the presence of the avalanche effect. This parameter was obtained for data at g$_{\mathrm{APD}}$ = 1 and discussed in the previous section \ref{sec:G_sys}. When higher gain levels are set, an additional multiplication gain M is achieved in the eAPDs by an electric field inside the detection region, which accelerates the primary photoelectrons and these strike out further (secondary) electrons in the multiplication region. Therefore, the resulting effective gain (G$\mathrm{_{eff}}$) must be higher than the G$\mathrm{_{sys}}$, for higher eAPD gains.

\begin{figure}[htb]
	\centerline{\includegraphics[width=0.5\textwidth]{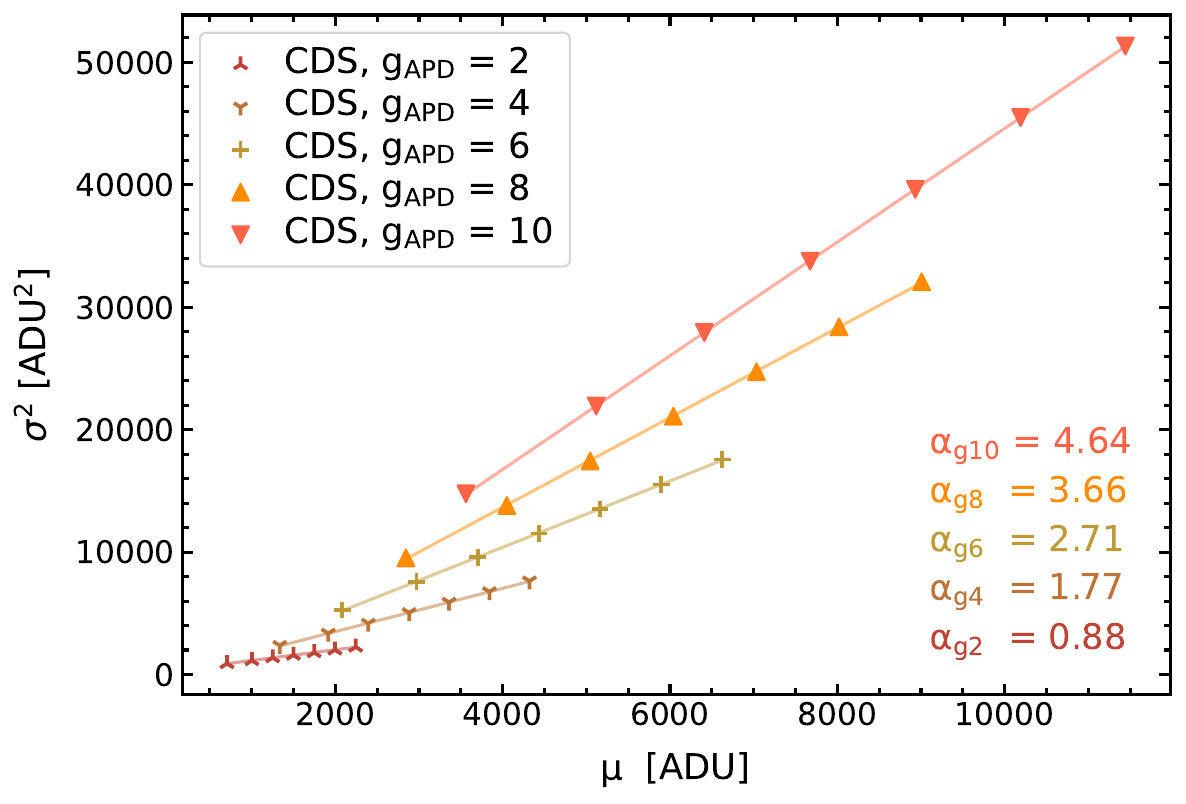}}
	\caption{PTCs of the correlated double sampling frames on avalanche gain levels 2, 4, 6, 8 and 10. The exposure times were changed from the minimal exposure up to 1.2 ms, in steps of 0.1 ms. A linear relation was fitted for the data on each avalanche gain level, with the corresponding slope parameters ($\mathrm{\alpha}$) shown in the lower right region of the plot. \label{fig:multiple_avalanche_gain_plot}}
\end{figure}

In Figure \ref{fig:multiple_avalanche_gain_plot} the photon transfer curve was constructed for each gain level together with the linear fit relations. We observe a linear behaviour in each set of points. For avalanche levels above g$_{\mathrm{APD}}$ = 2, the y-axis intercept of every fit assumes a ne\-ga\-tive value. The observed trend was not further investigated in this study.

\begin{figure}[t]
	\centerline{\includegraphics[width=0.5\textwidth]{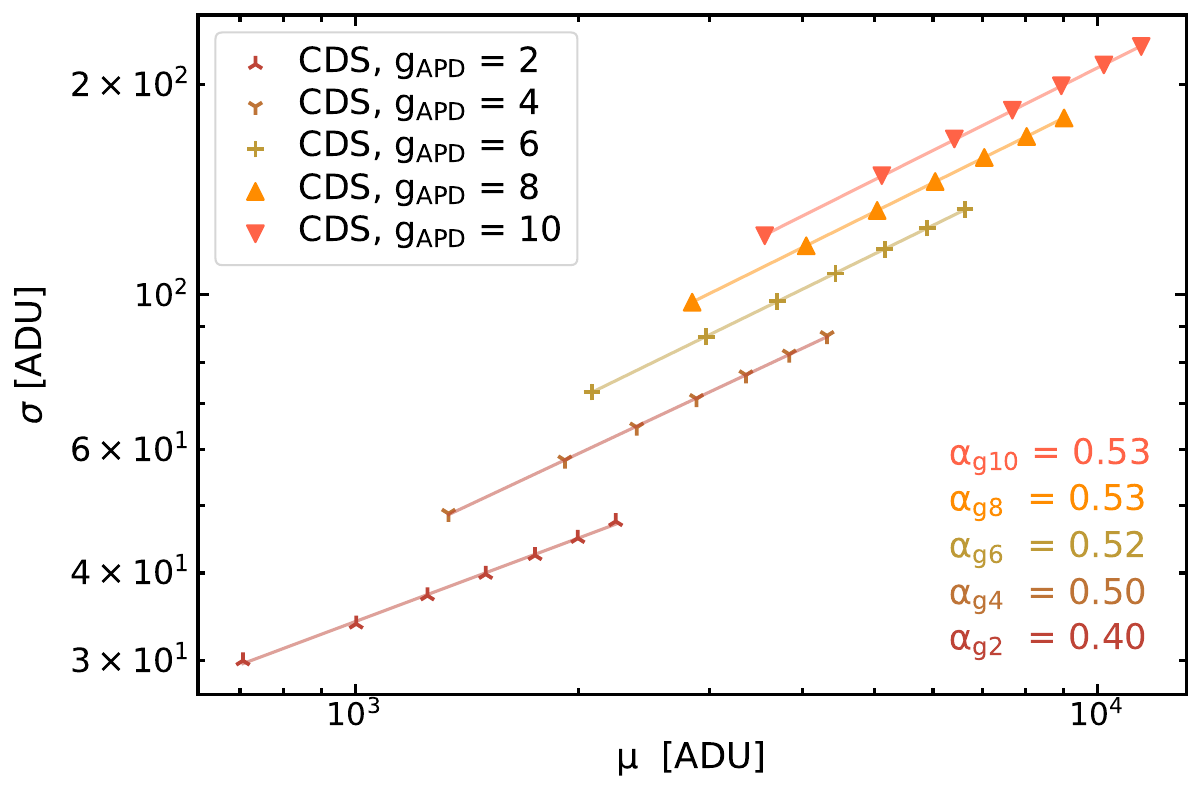}}
	\caption{Standard deviation versus average effective signal, both on a log-scale. The slope of the regression line drawn through each avalanche gain level data is shown in the lower right region of the plot.\label{fig:PTC_log-log_check}}
\end{figure}

We make some additional steps to verify the sanity of our data. As demonstrated in \cite{Janesick2001}, a PTC $\mathrm{\sigma-\mu}$ logarithmic graph of a well-behaved detector system should have a slope of 0.5 in the shot-noise regime. This is exactly what we can see in our data, when the same is presented as in the \hbox{Figure \ref{fig:PTC_log-log_check} }. 

A slope factor of $\sim$ 0.5 is observed for data at every gain level. Since the same exposure times were used for each set of data points, this resulted in a small deviation in the slope factor for lower gain level 2, where the signal acquired merely reached the shot-noise regime of the PTC. However, this will not play a significant role in the upcoming effective system gain calculation. 

\begin{figure}[ht]
	\centerline{\includegraphics[width=0.5\textwidth]{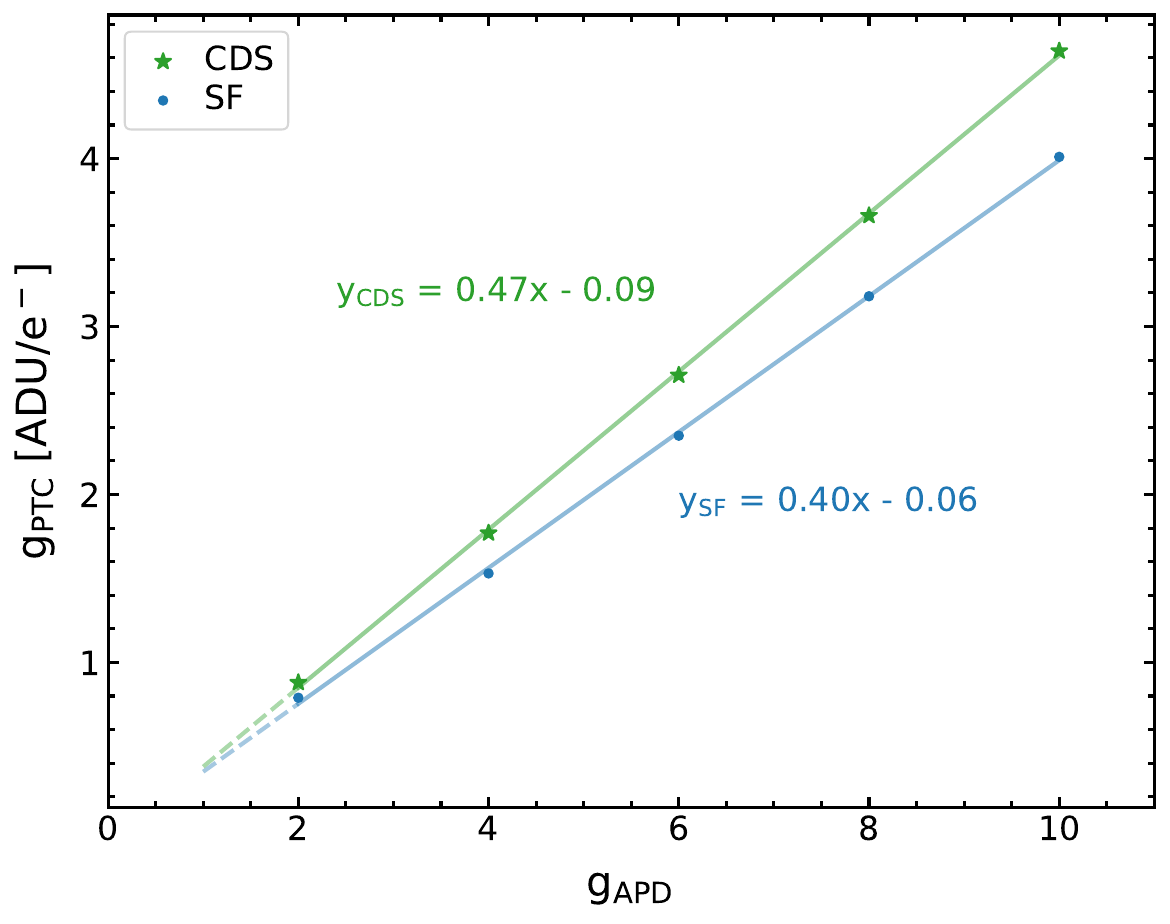}}
	\caption{Effective system gain (G$_{\mathrm{eff}}$) relation represented by the fit of PTC slope factors (g$_{\mathrm{PTC}}$) and the avalanche photodiode gain (g$_{\mathrm{APD}}$) levels, excluding g$_{\mathrm{APD}}$ = 1 data. A slight difference between the gain factors of CDS and of SF readouts lies in mapping their PTCs at different effective signal levels. \label{fig:slopes_of_slopes}}
\end{figure}

We start the procedure by taking the slopes from linear fits of data and comparing them to their associated eAPD gain levels. This is done for both CDS mode and single frame (SF) mode. The new relation is shown in Figure \ref{fig:slopes_of_slopes} both for data acquired in CDS and the SF mode.

When a new fit is applied to this data, we notice that the system exhibits a linear behaviour with the increasing gain.
Assuming this interrelation of gains is linear in every PTC regime (i.e. read, shot or fixed pattern noise dominated regime), we could use it to compute the real effective gain at any eAPD gain level, without the need of acquiring a new PTC diagram.

\subsection{Readout noise} \label{sec:RoN}

The readout noise is estimated as the Y interception of the fit on a PTC diagram, for avalanche gain = 1 data. From the fit (see Figure \ref{fig:gain_1}) we measured a variance of \hbox{242.65 ADU$^2$} and therefore the readout noise is $\approx$ 15.58 ADU. Applying the system gain, we convert this to approximately 33 e$^-$, a slightly lower value than 35 e$^-$ reported in \cite{Lanthermann2019} that used the same PTC analysis. 

In general, the readout noise ($\mathrm{\sigma_R}$) can also be described as the amount of variation the detector has between identical readouts. To measure the readout noise of the C-RED One on higher gain levels, we took one thousand dark single full-frames per each gain value, all at minimal exposure time. Calculating the standard deviation of every pixel over the multiple identical exposures allowed us to construct a standard deviation map or as we can call it - a "noise map". Median value of all pixels in this map represents a readout noise value (in ADU) of frames obtained at minimal exposure time and specific avalanche gain g$_{\mathrm{APD}}$. This is an adequate measure of readout noise, as the dark current and glow are sufficiently low at minimal integration time. In Table \ref{tab2} one can compare noise values of several of these maps. When expressed in analog-to-digital units, as in the third column of Table \ref{tab2}, we notice a very similar median noise value at g$_{\mathrm{APD}}$ = 1 and g$_{\mathrm{APD}}$ = 50. Naturally, since the signal at high g$_{\mathrm{APD}}$ = 50 is amplified as opposed to g$_{\mathrm{APD}}$ = 1 level, this means that a "true" noise value is much lower. To convert this value to e$^-$ units, it needs to be multiplied by the inverse of the appropriate effective system gain that can be found in the second column of the same table. The resulting readout noise in e$^-$ units is listed in the fourth column.   
\begin{center}
\begin{table}[htb]%
\caption{Results of the effective gain and the readout noise calculation for CDS frames at higher eAPD gains.\label{tab2}}
\centering
\begin{tabular*}{19.9pc}{@{\extracolsep\fill}lccc@{\extracolsep\fill}}
\toprule
\textbf{g$\mathrm{_{APD}}$} & \textbf{G$\mathrm{_{eff} \left[\frac{ADU}{e^-}\right]}$} \tnote{$^*$}  & \textbf{$\mathrm{\sigma_{g_x}[ADU]}$}  & \textbf{$\mathrm{\sigma_R[e^-]}$} \tnote{$^{**}$}  \\
\midrule
10  & 4.61  & 16.681  & 3.62   \\
20  & 9.31  & 16.774  & 1.80   \\
30  & 14.01  & 16.909  & 1.21  \\
40  & 18.71  & 17.097  & 0.91   \\
50  & 23.41  & 17.325  & 0.74   \\
\bottomrule
\end{tabular*}   
\begin{tablenotes}
\item[\hspace{0.15cm}$^* $] \hspace{5pt} G$\mathrm{_{eff} = 0.47 \cdot g_{APD} - 0.09}$ 
\item[\hspace{0.15cm}$^{**} $] \hspace{1pt} $\mathrm{\sigma_R = \sigma_{g_x}\cdot (G\mathrm{_{eff}})^{-1}}$
\end{tablenotes}
\end{table}
\end{center}

\begin{figure}[ht]
	\centerline{\includegraphics[width=0.5\textwidth]{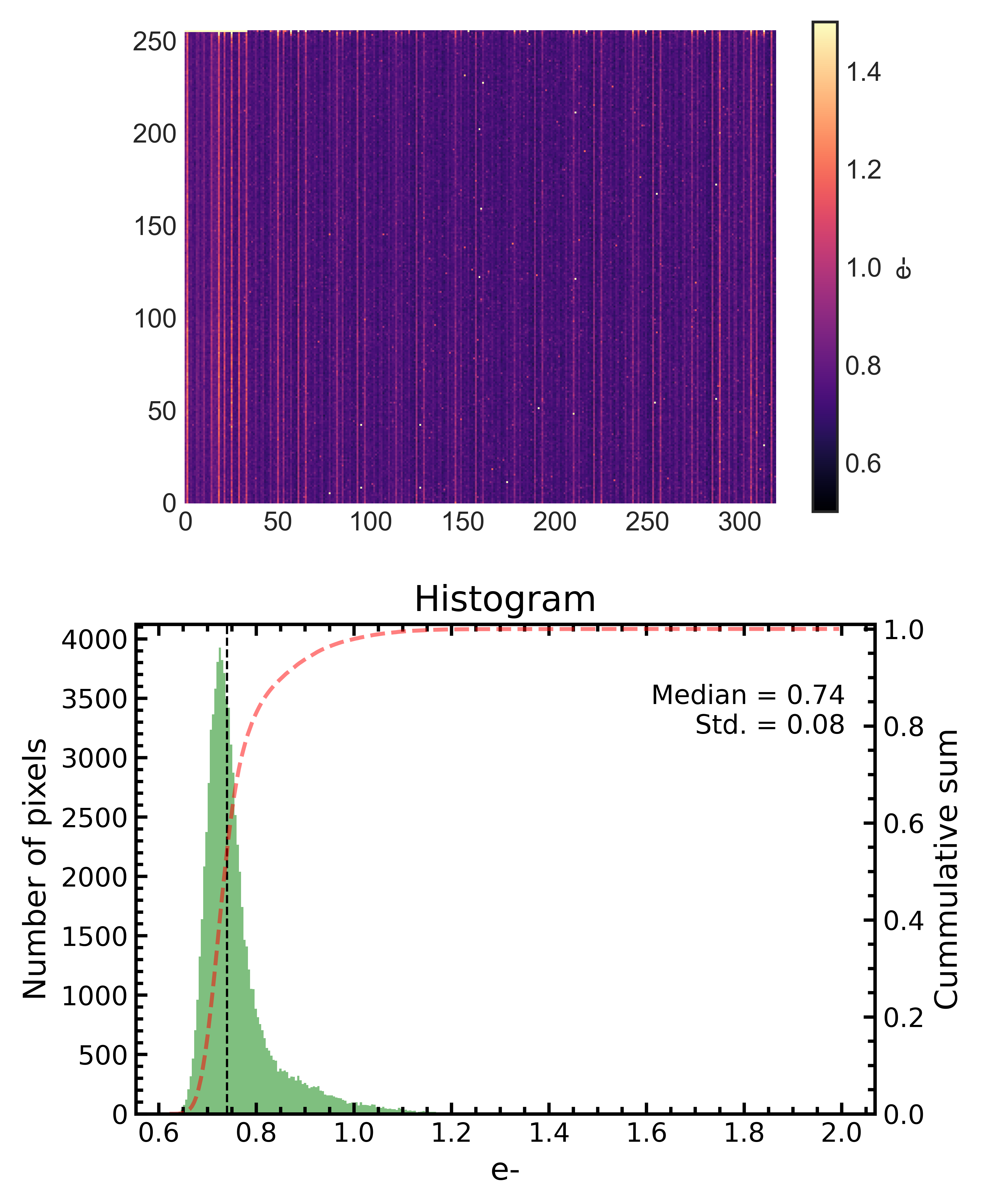}}
	\caption{Noise map and histogram at avalanche gain setting 50. \label{fig:noise_hist_sube}}
\end{figure}

Although \cite{Feautrier-cred1-cred2} states that a camera is reaching subelectron readout noise values for eAPD gains \hbox{> 30}, we report 1.21 e$^-$ at the same gain level. From g$_{\mathrm{APD}}$ = 40 we measure camera entering the subelectron regime. \hbox{At g$_{\mathrm{APD}}$ = 50} it reaches a value of 0.74 e$^-$, comparable to the value of \hbox{0.67 e$^-$} that was reported in \cite{Feautrier2022} for the same type of frames and setup (CDS mode, \hbox{gain 50}, minimal exposure, room temperature source). Figure \ref{fig:noise_hist_sube} shows the noise map and the corresponding histogram for avalanche gain 50. The analysis of the histogram showed that only 0.37\% of the pixels are above a 5 $\times$ sigma noise level, which indicates a good overall pixel performance.

\subsection{Dynamic range}

\begin{figure}[h!]
	\centerline{\includegraphics[width=0.5\textwidth]{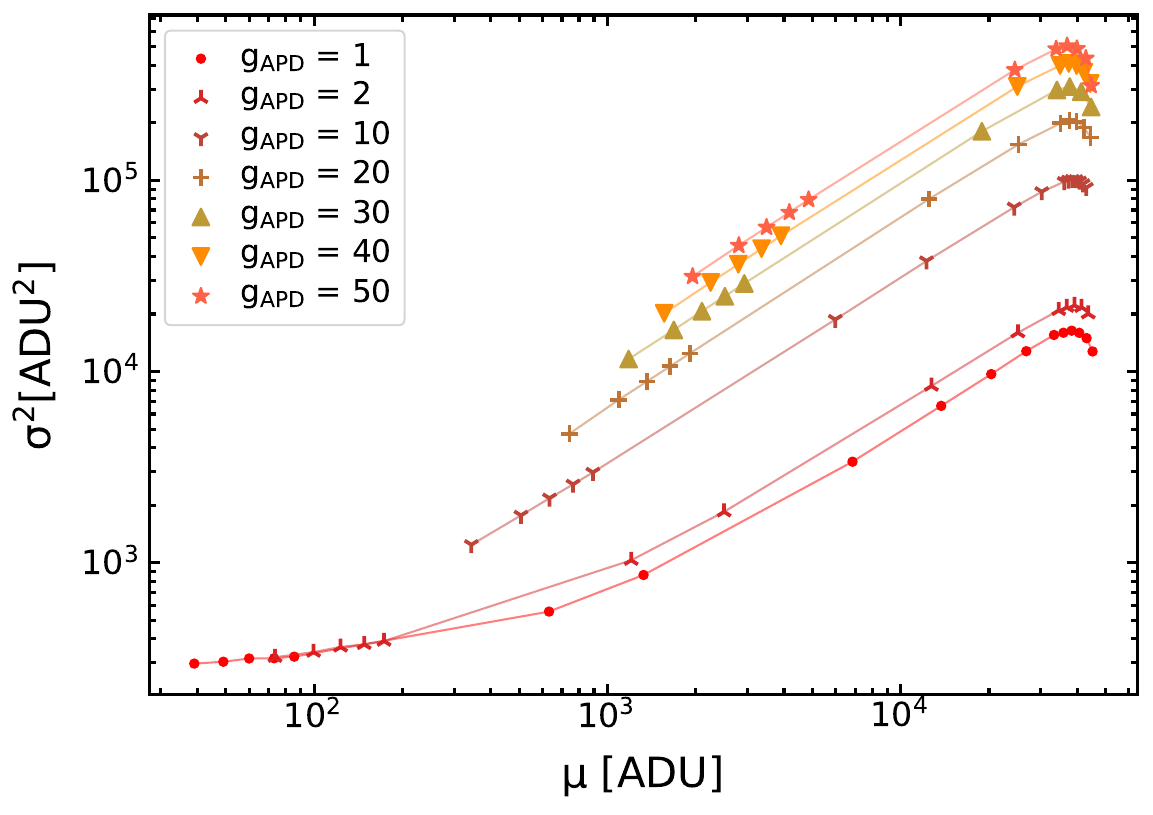}}
	\caption{PTCs obtained from data on different APD gain levels, in a wide range of exposure times. Values are expressed in analog-to-digital units.\label{PTC_dynamic_range_ADUs}}
\end{figure}

\begin{figure}[h!]
	\centerline{\includegraphics[width=0.5\textwidth]{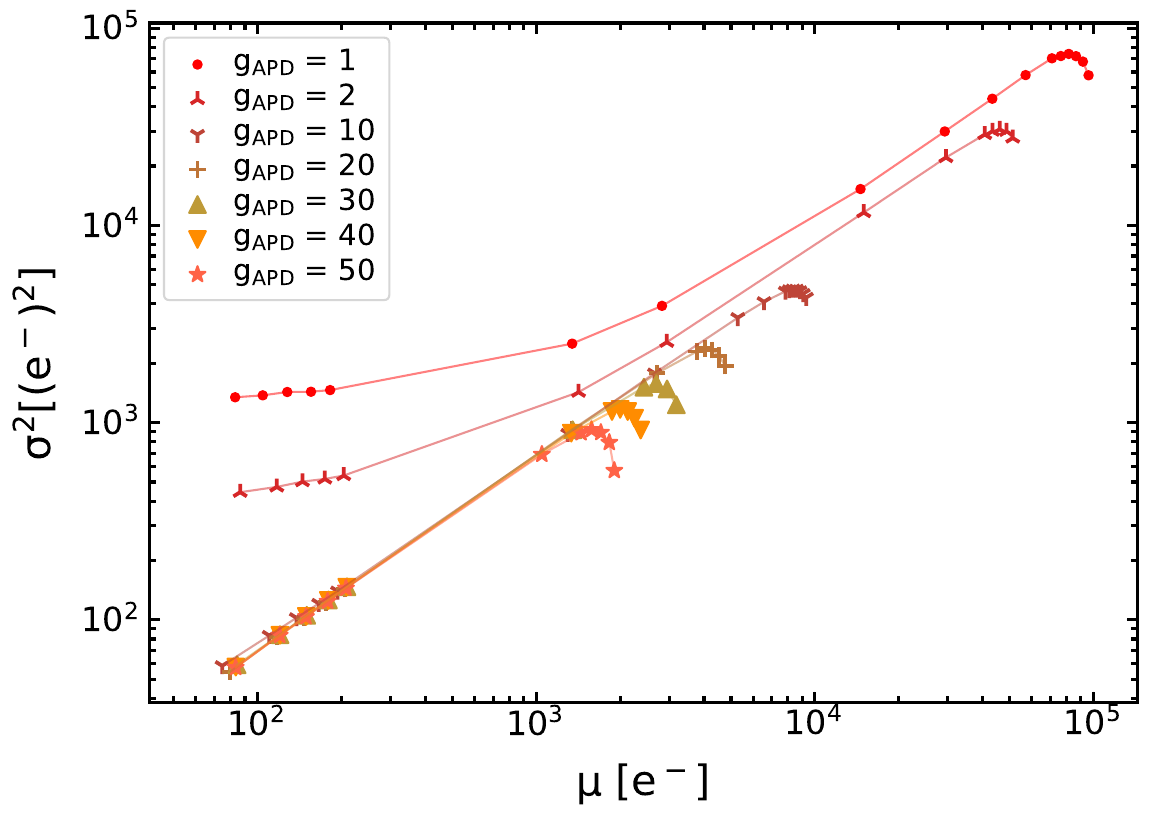}}
	\caption{PTCs obtained from data on different APD gain levels, in a wide range of exposure times. Values are expressed in electron units. We reach saturation once the variance of the noise starts declining.\label{PTC_dynamic_range_e-}}
\end{figure}

According to ~\cite{Janesick2001} the dynamic range (DR) can be understood as the ratio between brightest and faintest objects that can be simultaneously differentiated. Signal of the brightest objects would be at the near-saturation level of an image sensor, while the faintest objects would be comparable to the dark noise level of the imager. Therefore, dynamic range can be defined in the following way:
\begin{equation}
\mathrm{DR = \frac{\mu_{fw}(e^-)}{\sigma_R(e^-)},} 
\label{eq1}
\end{equation}
where $\mathrm{\mu_{fw}(e^-)}$ can be described as the full well capacity and $\mathrm{\sigma_R(e^-)}$ the read noise parameter. The full-well is a measure of the amount of charge a pixel can hold. When the well capacity is filled, the measured variance on the PTC decreases. This can be well observed in Figure \ref{PTC_dynamic_range_ADUs} and \ref{PTC_dynamic_range_e-} where the noise value suddenly decreases on the right end of PTCs. Data is plotted on a log-log scale in order to cover the whole dynamic range of the camera. 

As can be observed from Figure \ref{PTC_dynamic_range_e-}, through the avalanche process less signal electrons are needed to reach the full well value. Knowing the readout noise values at several different g$_{\mathrm{APD}}$ levels (see previous section \ref{sec:RoN}) and estimating the full well capacity in e$^-$ units we are able to examine how DR varies with increasing g$_{\mathrm{APD}}$. The DR can also be expressed in decibels (dB) with 
\begin{equation}
\mathrm{DR = 20 \log \left(\frac{\mu\mathrm{_{fw}}(e^-)}{\sigma_\mathrm{R}(e^-)}\right)},
\label{eq2}
\end{equation}

\smallskip

\begin{figure}[t]
	\centerline{\includegraphics[width=0.5\textwidth]{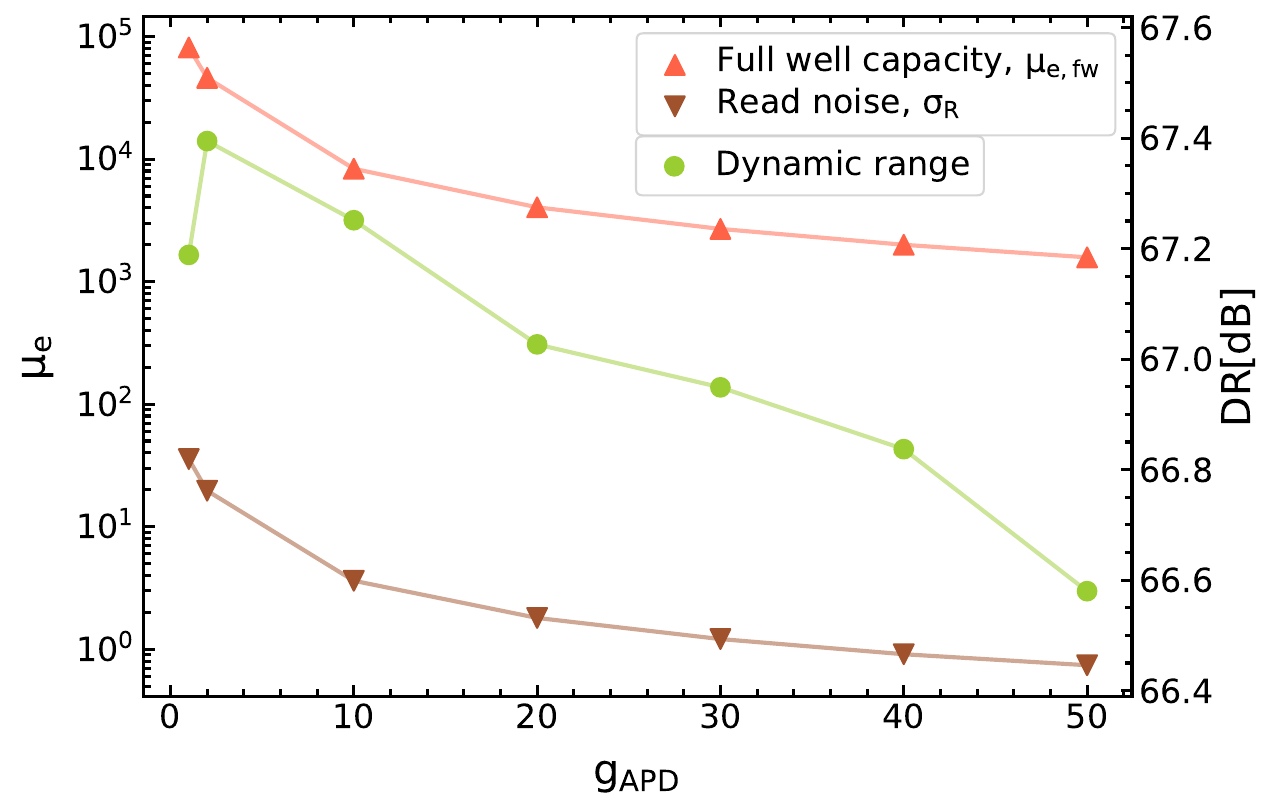}}
	\caption{Comparison of full well capacity and read noise thresholds with the dynamic range (DR) being computed as the ratio of the former parameter versus the latter one. Left vertical axis shows the mean signal in e$^-$ units, right vertical axis shows the ratio value of DR calculation and horizontal axis contains a list of avalanche gain values.  \label{fig6}}
\end{figure}

This is presented in Figure \ref{fig6}. Using no avalanche gain, i.e. \hbox{g$_{\mathrm{APD}}$ = 1}, the resulting dynamic range value is equal to \hbox{67.19 dB}. At g$_{\mathrm{APD}}$ = 50, the resulting dynamic range equals \hbox{66.58 dB}. Although fewer charge is needed to reach the full-well at higher avalanche settings, the detector is able to operate at lower flux levels, which accounts for a relative stable dynamic range during the analyzed avalanche gain range resulting in a decline lower than 1 dB. 

Although the manufacturer has not specified DR value for the C-RED One camera, one can compare the resulting value to those provided for C-RED 2, another FLI's high-frame rate NIR camera. In the basic operational mode, C-RED 2 possess 63 dB. Therefore, C-RED One has excellent performances even at higher gain values, where it retains a rather good dynamic range properties.

\section{Conclusion}
\label{sec:conclusion}

We have presented a simple and unique method to determine effective system gain without the need to construct a PTC diagram at each eAPD gain of interest. The importance of this method consists in the fact that once the effective gain relation is found, it can be used for data acquired in different flux regimes, bypassing a possible non-linearity of the camera system. 

We also performed a classical photon transfer method analysis in order to determine the basic system gain and readout noise at gain = 1. To determine readout noise of higher eAPD gain data, we used appropriate effective system gain factors. The readout noise was reduced to a minimum of 0.74 e$^-$ at \hbox{gain = 50}. The results presented here are consistent with other literature sources and those in the test report we received with our C-RED One. 

Although an extensive characterization of the SAPHIRA detector and the C-RED One camera already exists, it can still be complemented by the effective gain analysis and extended with the dynamical range analysis provided in this paper. Since both parameters are very important in cases where one operates in minimal light conditions, a correct estimation of these parameters helps us in understanding the limitations of our instruments and the expected behaviour once it is integrated in a complete experimental setup. 


\section*{Acknowledgments}

This work was supported by \fundingAgency{Bundesministerium für Bildung und Forschung} under Contract No. \fundingNumber{03Z22AB1A NIR-DETECT}, \fundingNumber{03Z22AN11 Astrophotonics}, \fundingNumber{03Z22A51 Meta-Zik AstrOOPtics}, and \fundingNumber{03Z22AI1 Strategic Investment}. We thank the staff from the Technical Section at AIP for all their support. We are also grateful to the anonymous referee for the comments which led to a number of improvements to this paper.



\bibliography{Wiley-ASNA}%

\end{document}